\documentclass[12pt]{article}

\usepackage{amsmath,amssymb,latexsym,cite}
\usepackage[]{hyperref}

\newcommand\coolleftbrace[2]{%

#1\left\{\vphantom{\begin{matrix} #2 \end{matrix}}\right.}

\begin{document} 
\begin{center}

\vspace*{0.5in}

{\large\bf Toda-like (0,2) mirrors to products of projective spaces}

\vspace{0.2in}

Zhuo Chen$^1$, Eric Sharpe$^1$, Ruoxu Wu$^1$

\vspace*{0.2in}

\begin{tabular}{l}
$^1$ Physics Department \\
Robeson Hall (0435) \\
Virginia Tech \\
Blacksburg, VA  24061 
\end{tabular}

{\tt zhuo2012@vt.edu}, {\tt ersharpe@vt.edu}, {\tt ronwu@vt.edu}

$\,$

\end{center}

One of the open problems in understanding (0,2) mirror symmetry
concerns the construction of 
Toda-like Landau-Ginzburg mirrors to (0,2) theories on Fano spaces.
In this paper, we begin to fill this gap by
making an ansatz for (0,2) Toda-like theories mirror to (0,2) supersymmetric
nonlinear sigma models on products of projective spaces, with deformations
of the tangent bundle, generalizing a special case previously worked out
for ${\mathbb P}^1 \times {\mathbb P}^1$.  We check this ansatz
by matching correlation functions
of the B/2-twisted Toda-like theories to correlation functions of
corresponding A/2-twisted nonlinear sigma models, computed primarily
using localization techniques.
These (0,2) Landau-Ginzburg models admit redundancies, which can lend 
themselves to multiple distinct-looking representatives of the
same physics, which we discuss.

\begin{flushleft}
March 2016
\end{flushleft}

\newpage

\tableofcontents

\newpage

\section{Introduction}

Historically, mirror symmetry has been one of the most productive arenas for
mathematics to emerge from string theory.  It has led to notions of curve
counting, quantum cohomology, and Gromov-Witten theory, and has been
generalized via {\it e.g.} homological mirror symmetry \cite{kont-hms}.

This paper concerns a different generalization of mirror symmetry, known as
`(0,2) mirror symmetry,' as it relates UV descriptions of
theories with (0,2) supersymmetry,
just as ordinary mirror symmetry relates UV descriptions of theories
with (2,2) supersymmetry.

Although (0,2) mirror symmetry has not been developed to nearly the
same extent as ordinary mirror symmetry, a number of crucial results do exist.
One of the first accomplishments was a numerical scan through
anomaly-free examples demonstrating the existence of pairs of (0,2)
theories with matching spectrum computations \cite{bsw}, giving strong
evidence for the existence of (0,2) mirrors.  Other work includes
a version \cite{blum-sethi} of the old Greene-Plesser
orbifold construction \cite{gp}, work on GLSM-based dualities
\cite{0309226}, and most recently, a proposal for a generalization of
Batyrev's construction involving reflexively plain polytopes
\cite{mel-ron-mirror}.  In addition, there has been considerable work
on quantum sheaf cohomology 
\cite{0406226,0502064,0605005,0604179,0705.0790,0710.2354,0712.1058,0712.3272,0801.3955,08100012,0902.3908,1001.2104,1010.4667,1101.1305,1103.1322,1110.3751,1110.3752,esbrazil,dlm,1512.08058,1512.08586}, 
the (0,2) analogue of
ordinary quantum cohomology. 

All that said, many basic gaps remain.  
For example, there is not yet
a systematic description of (0,2) Landau-Ginzburg mirrors to
(0,2) nonlinear sigma models on
Fano spaces, aside from a special case discussed in \cite{0309226}.  
This paper is a first pass at filling that gap.

Recall a (0,2) supersymmetric nonlinear sigma model is typically defined by
a complex K\"ahler manifold $X$ and holomorphic vector bundle
${\cal E} \rightarrow X$ obeying
\begin{displaymath}
{\rm ch}_2({\cal E}) \: = \: {\rm ch}_2(TX),
\end{displaymath}
known as the Green-Schwarz or anomaly cancellation condition.
In addition, to define the A/2-twist, we must also require that
\begin{displaymath}
\det {\cal E}^* \: \cong \: K_X.
\end{displaymath}
For example, if ${\cal E} = TX$, both of these conditions are trivially
satisfied.  There is also a B/2-twist, which requires instead
\begin{displaymath}
\det {\cal E} \: \cong \: K_X.
\end{displaymath}
If ${\cal E} = TX$ and $K_X^{\otimes 2}$ is trivial, these conditions
are satisfied, which match the conditions for consistency of the
closed-string B model \cite{0605005}.
The A/2 and B/2 twists are closely related:  the A/2 twist of
a nonlinear sigma model defined by $(X, {\cal E})$ is equivalent 
to the B/2 twist of a nonlinear sigma model defined by
$(X, {\cal E}^*)$ \cite{0605005}.

For $X$ a Calabi-Yau, the simplest version of (0,2) mirror symmetry
asserts that the pair $(X, {\cal E})$ define the same (0,2) SCFT
as another pair $(X', {\cal E}')$, satisfying the same two conditions
above, where $X'$ is Calabi-Yau.  This duality also exchanges the
A/2 and B/2 twists, in the sense that the A/2 twist of the nonlinear
sigma model defined by $(X, {\cal E})$ is equivalent to the B/2 twist
of the nonlinear sigma model defined by $(X', {\cal E}')$.

In this paper, we will be concerned
with duals in cases where $X$ is not Calabi-Yau.  Specifically, we will
consider duals to A/2 twists of nonlinear sigma models on Fano manifolds
$X$, which will correspond to B/2 twists of certain (0,2) Landau-Ginzburg
models. 

For (2,2) theories, such dualities are well-known as Toda duals to Fano
spaces.  For (0,2) theories, one special case was worked out in
\cite{0309226}, corresponding to particular deformations of the tangent
bundle of ${\mathbb P}^1 \times {\mathbb P}^1$.  The point of this paper is
to construct (0,2) Landau-Ginzburg mirrors to more general tangent
bundle deformations of arbitrary products of projective spaces, 
as deformations of (2,2) Landau-Ginzburg mirrors, and in so
doing, pave the way for an understanding of such duals to arbitrary
Fano manifolds.

We check our ansatz for (0,2) duals by comparing correlation
functions of B/2 twists of the proposed (0,2) Landau-Ginzburg mirrors
to correlation functions in A/2-twisted nonlinear sigma models,
which can be computed as in \cite{08100012,1110.3751,1110.3752,1512.08058}.  
In particular, those
nonlinear sigma models compute quantum sheaf cohomology, a generalization
of ordinary quantum cohomology.  Recall that in a (2,2) supersymmetric
nonlinear sigma model, the ordinary quantum cohomology is generated
additively by
\[
H^{\bullet}(X, \wedge^{\bullet}T^*X),\]
with $T^*X$ the cotangent bundle of $X$.   In the (0,2) case,
the analogue (known as the quantum sheaf cohomology ring) is generated
additively by
\[
H^{\bullet}(X, \wedge^{\bullet}{\cal E}^*)\]
instead. Quantum sheaf cohomology was first introduced in \cite{0406226}, 
and the subject has been further developed in a number of works including 
\cite{1110.3751,1101.1305,1110.3752,0710.2354,1001.2104,1010.4667, 
0712.3272,08100012,1103.1322,0902.3908,0712.1058,0502064,0605005, 
0604179,0705.0790,0801.3955,1512.08058,1512.08586,esbrazil,dlm}.

We will see in sections \ref{p1}, \ref{pn}, that
the (0,2) Toda-like duals have the property that their classical vacua
are defined by the quantum sheaf cohomology ring relations of the
dual A/2-twisted theories.

In section \ref{sec22}, we begin by reviewing old results from 
ordinary mirror symmetry. In section \ref{p1}, we describe our ansatz
for the (0,2) Toda-like dual to ${\mathbb P}^1 \times {\mathbb P}^1$
with a general deformation of the tangent bundle, and check that
(genus zero) correlation functions match those of the
corresponding A/2 theory.  We then compare
the number of parameters in the theory to the number of expected
infinitesimal moduli, and discuss some reparametrization symmetries that
can be used to write physically-equivalent but different-looking forms
of the Toda-like dual. 
In section \ref{pn}, we generalize to products of projective spaces,
discussing Toda-like duals, giving a general argument for matching of
(genus zero) correlation functions to those of the corresponding A/2 theory,
and also checking in detail in the special case of
${\mathbb P}^1 \times {\mathbb P}^2$.  In appendices, we give
detailed results for correlation functions in a number of examples,
and also discuss the number of moduli appearing mathematically..

\section{Review of Toda models in ordinary mirror symmetry}\label{sec22}

Let us quickly review ordinary Toda duals to A-twisted (2,2) supersymmetric
nonlinear sigma models on projective spaces.
First, recall that in the A-twisted\footnote{
The reader should note that we do not couple this theory to
worldsheet gravity -- throughout this paper, we consider only
topological field theories, not topological string theories.
} nonlinear sigma model on ${\mathbb P}^n$, all BRST-cohomology classes
of local operators are generated by a single operator $\psi$, corresponding
to a degree-two cohomology class on ${\mathbb P}^n$, with correlation functions
of the form
\begin{eqnarray*}
\langle \psi^n \rangle  & = & 1, \\
\langle \psi^{2n+1} \rangle & = & q, \\
\langle \psi^{n+d(n+1)} \rangle & = & q^d,
\end{eqnarray*}
and OPE (quantum cohomology relation) $\psi^{n+1} = q$.

The mirror Toda theory 
is a B-twisted Landau-Ginzburg theory with superpotential
of the form
\[
W=\exp(Y_1)+\exp(Y_2)+\cdots+\exp(Y_n) + q\exp(-Y_1-Y_2-\cdots-Y_n).\]
(In effect, because of the exponentials, the superpotential is
defined over $({\mathbb C}^{\times})^n$.)  We define
\begin{equation*}
X_i=e^{Y_i} ,
\end{equation*}
so that the superpotential can be written in the simpler form
\begin{displaymath}
W \: = \: X_1 + X_2 + \cdots + X_n + \frac{q}{X_1 \cdots X_n},
\end{displaymath}
bearing in mind that the fundamental fields are $Y_i$.

As the superpotential is over a vector space,
the correlation functions in this\footnote{
Here we are considering Landau-Ginzburg models over vector spaces,
for which this correlation function can be found in
\cite{vafa-toplg}.
See \cite{0801.3836} for a discussion of correlation functions in
more general B-twisted Landau-Ginzburg models.  The computation in
this section, demonstrating how quantum cohomology appears in 
Toda duals, is also described in \cite{0502053}, 
as a prelude to the discussion of
Toda duals to (2,2) theories on smooth Fano Deligne-Mumford stacks.
} theory are
\[
\langle F_1\cdots F_n\rangle=\sum\limits_{dW=0}^{}\frac{F_1\cdots F_n}{H},\]
where $H=\det(\partial_i\partial_jW)$ (with derivatives computed with
respect to $Y$'s).

Solving the constraint $dW=0$ (for derivatives
with respect to the fundamental fields $Y$), one finds that the
classical vacua are given by
\[
X_1=X_2=\cdots=X_n\equiv X, \: \: \:
X = q X^{-n}. \]
In particular, the vacua are given by $X$ such that
$X^{n+1} = q$, which is the defining relation
of the quantum cohomology ring of ${\mathbb P}^n$.
(This is no accident, and in fact, is an important property we will
apply later in working out duals to (0,2) theories.)
Furthermore, after restriction to the classical vacua,
the Hessian $H$ is easily computed to be
\begin{displaymath}
H \: = \: (n+1) X^n.
\end{displaymath}
Thus, the correlation functions of this model are\[
\langle X^m\rangle=\sum\frac{X^m}{(n+1)X^n},\]
where the sum runs over $X$'s solving $X^{n+1}=q$, {\it i.e.}
$(n+1)$th roots of $q$.
This expression can only be nonvanishing when $m-n$ is divisible by $n+1$. 
We find that the nonzero correlation functions are\footnote{
In principle we should write the correlation functions in terms of
$X_1, \cdots, X_n$; however, since the $X_i$ coincide on the set of vacua,
and the correlation functions are computed by summing over vacua, it is
an immediate result that
\begin{displaymath}
\langle f(X_1, \cdots, X_n) \rangle \: = \: \langle f(X, \cdots, X) \rangle,
\end{displaymath}
so without loss of generality
we merely write the correlation functions in terms of powers of $X$.
}
\[\begin{array}{rcl}
\langle X^{n}\rangle&=& 1,\\
\langle X^{2n+1}\rangle&=& q,\\
\langle X^{n+d(n+1)}\rangle&=& q^d,
\end{array}\]
matching the A model correlation functions if we identify $X$ with
$\psi$.

In the rest of this paper, we shall describe an ansatz for Toda-like
duals to (0,2) nonlinear sigma models on certain
Fano spaces with deformations
of the tangent bundle, generalizing the discussion above off
the (2,2) locus, which we will check by comparing correlation
functions (and quantum sheaf cohomology relations).

\section{Toda-like duals to $\mathbb{P}^1\times\mathbb{P}^1$}\label{p1}

\subsection{The (0,2) NLSM}
\label{sect:p1p1-a2}

In the case of $X=\mathbb{P}^1\times\mathbb{P}^1$, one can describe a general
deformation $\cal E$ of the tangent bundle as the cokernel of the following 
sequence:\[
0 \: \longrightarrow \: {\cal O}\otimes{\cal O} \: \xrightarrow{E} \: 
{\cal O}(1,0)^2\oplus{\cal O}(0,1)^2 \: \longrightarrow \: 
{\cal E} \: \longrightarrow \: 0 ,
\]
where 
$$E=\left[\begin{array}{lr}
Ax & Bx\\
C\tilde{x} & D\tilde{x}\\
\end{array}\right],$$ 
with $A$, $B$, $C$, $D$ $2\times 2$ matrices and \[
x=\left[\begin{array}{c}
x_1\\
x_2\\
\end{array}\right], \qquad \tilde{x}=\left[\begin{array}{c}
\tilde{x}_1\\
\tilde{x}_2\\
\end{array}\right]\]
are homogeneous coordinates on the two $\mathbb{P}^1$ factors. 
The tangent bundle corresponds to $A=D=I$, and $B=C=0$. 
For more general $A$, $B$, $C$, $D$, the vector bundle is (generically)
a deformation of the tangent bundle. 
In this model, it has been argued in 
\cite{08100012,1110.3751,1110.3752,1512.08058} that the OPE ring
relations in the A/2 twist (defining the quantum sheaf cohomology ring)
are given by
\begin{align}
\det(A\psi+B\tilde{\psi})&=q_1,\\
\det(C\psi+D\tilde{\psi})&=q_2.
\end{align}

Correlation functions in A/2 twisted theories on ${\mathbb P}^1 \times
{\mathbb P}^1$ with a deformation of the tangent bundle can be computed
in several ways.  One method is to use direct Cech techniques to compute
sheaf cohomology products on ${\mathbb P}^1 \times {\mathbb P}^1$,
as has been discussed in {\it e.g.} \cite{0406226,0710.2354,esbrazil}.
Another method is to use GLSM-based Coulomb branch results, as described
in \cite{08100012}.  A third, more recent, method is to use residue
formulas obtrained via localization, as in \cite{1512.08058}.
In this last approach, correlation functions in the A/2 twisted theory on
${\mathbb P}^1 \times {\mathbb P}^1$ are of the form\footnote{
As a matter of
principle, there is a phase ambiguity in expressions of this
form, due geometrically to possible phases of the isomorphism $\det {\cal E}^*
\stackrel{\sim}{\longrightarrow} K_X$, and physically to chiral
left and right global $U(1)$ actions on the worldsheet, that play a role
closely analogous to that of the Bagger-Witten line bundle.
The expression given here implicitly determines such phases.
}
\begin{eqnarray*}
\lefteqn{
\langle f(\psi,\tilde{\psi}) \rangle
} \\
&  = &
\sum_{k_1, k_2} q_1^{k_1} q_2^{k_2} {\rm JKG-Res}\left(
\frac{1}{ \det(A \psi + B \tilde{\psi})^{k_1+1} }
\frac{1}{\det(C \psi + D \tilde{\psi})^{k_2+1} }
f(\psi, \tilde{\psi} ) \right).
\end{eqnarray*}

However one computes the correlation functions, the results have the
following form, in terms of the matrices $A$, $B$, $C$, $D$ above.
Let
\begin{displaymath}
a=\det(A), b=\det(B), c=\det(C), d=\det(D),
\end{displaymath}
\begin{displaymath}
e=\det(A+B),\quad f=\det(C+D) .
\end{displaymath}
Define
\begin{displaymath}
\mu = e - a - b, \: \: \:
\nu = f - c - d,
\end{displaymath}
\begin{eqnarray*}
\phi_1&=& \nu b - \mu d \: = \: ad+bf-de-bc,\\
\phi_2&=& ad-bc,\\
\phi_3&=& \mu c - \nu a \: = \: ad+ce-af-bc,
\end{eqnarray*}
\begin{eqnarray*}
\Delta & = & \phi_2^2 - \phi_1 \phi_3, \\
& =& (c-d)(bc-ad)e+cde^2+(a-b)(ad-bc)f-(bc+ad)ef+abf^2.
\end{eqnarray*}
Then the two-point correlation functions, for example, can be expressed as:
\begin{equation}
\langle \psi\psi\rangle = \frac{\phi_1}{\Delta}, \: \: \: 
\langle \psi\tilde{\psi}\rangle = \frac{\phi_2}{\Delta}, \: \: \: 
\langle \tilde{\psi}\tilde{\psi}\rangle = \frac{\phi_3}{\Delta}  \label{A1}.
\end{equation}

Higher-point correlation functions have a similar form.
We list four-point functions in this A/2-twisted theory in
appendix~\ref{app:p1p1-a2-results}.
More general correlation functions at genus zero are straightforward to
compute with residue techniques, but the resulting expressions are rather
unwieldy, so we do not include them in this paper.

\subsection{The Toda-like mirror theory}
\label{sect:p1p1lg}

We claim the 
mirror theory to the A/2 twisted theory just described,
is a (0,2) Landau-Ginzburg model, defined by
a (0,2) superpotential of the form
\begin{equation}
W=FJ+\tilde{F}\tilde{J},
\end{equation}
where $F$ and $\tilde{F}$ are Fermi superfields, and
\begin{align*}
J&=X^{-1}(\det(AX+B\tilde{X})-q_1)=aX+b\frac{\tilde{X}^2}{X}+\mu\tilde{X}-\frac{q_1}{X},\\
\tilde{J}&=\tilde{X}^{-1}(\det(CX+D\tilde{X})-q_2)=
d\tilde{X}+c\frac{X^2}{\tilde{X}}+\nu X-\frac{q_2}{\tilde{X}},
\end{align*}
for $X = \exp(Y)$, $\tilde{X} = \exp(\tilde{Y})$, where $Y$,
$\tilde{Y}$ are the fundamental fields, and
\begin{displaymath}
a=\det A, \quad b=\det B, \quad c=\det C, \quad d=\det D,
\end{displaymath}
\begin{eqnarray*}
\mu&=&\det(A+B)-\det A-\det B,\\
\nu&=&\det(C+D)-\det C-\det D,
\end{eqnarray*}
for $A$, $B$, $C$, $D$ the matrices defining the tangent bundle
deformation of the A/2 theory.

In passing, the form written here does not manifestly match the
expression in \cite{0309226} for the special case they considered.
In section~\ref{sect:redundancy}, 
we will study various field redefinitions yielding
non-obviously-equivalent expressions,
and discover the expression in \cite{0309226} arising as a special case.

We will check the ansatz above by comparing correlation functions between
the original A/2 theory and the B/2 twist of the Landau-Ginzburg theory above,
but first, let us make a few quick observations.

As one consistency check, note that for
\[A=D=I,\; B=C=0,\] 
then the vector bundle $\cal E$ is the tangent bundle, and the theory has
(2,2) supersymmetry. This also can be seen from the (0,2) superpotential
\[
W=F\left(X-\frac{q_1}{X}\right) +
\tilde{F}\left(\tilde{X}-\frac{q_2}{\tilde{X}}\right),
\]
which matches the (2,2) superpotential in this case.

As another check, note that
the space of classical vacua of this theory 
($J=\tilde{J}=0$) matches the space of solutions to the 
quantum sheaf coholomogy ring relations:\begin{align}
\det(AX+B\tilde{X})&=q_1,\\
\det(CX+D\tilde{X})&=q_2.
\end{align}

Now, let us compute and compare genus zero correlation functions.
Given a B/2-twisted Landau-Ginzburg model with superpotential $W$
over a vector space or a product of ${\mathbb C}^{\times}$'s, 
correlation functions at genus zero are given by\footnote{
Correlation functions for more general B/2-twisted Landau-Ginzburg models
are discussed in \cite{0801.3955}.  In passing, we should comment on the
absence of worldsheet instanton corrections to the formulas above.
On the (2,2) locus, the Toda duals to A model topological field theories
are B-twisted, and correlation functions in the B model do not have
worldsheet instanton corrections.  
In the present case, our Toda-like mirrors to A/2 model 
pseudo-topological field theories are B/2 twisted.  Unlike the (2,2) case,
however, in general B/2 twisted models can and will receive worldsheet
instanton corrections.

However, our Toda-like theories are defined by superpotentials over
algebraic tori, {\it i.e.} $({\mathbb C}^{\times})^n$,
and there are no non-constant holomorphic maps
from ${\mathbb P}^1$ (or any projective variety) to an 
algebraic torus.  All holomorphic maps are constant maps, 
hence there are no worldsheet instanton
corrections in these theories \cite{katzpriv}.  
Thus, we need only compute classically
in the B/2 model, just as in ordinary Toda mirrors.
}
\cite{0712.1058} \begin{equation}
\langle \phi^{i_1}(x_1)\cdots \phi^{i_k}(x_k) \rangle=\sum_{J_i(\phi)=0}^{}\phi^{i_1}(x_1)\cdots \phi^{i_k}(x_k)[\det\limits_{i,j}J_{i,j}]^{-1} \label{B/2 corr.}
\end{equation}
where the sum is over classical vacua.

Using the formula above for B/2-twisted Landau-Ginzburg correlation
functions, one finds that
the two-point correlation functions in this model are given by
\[\begin{array}{rcl}
\langle XX\rangle&=&\Delta^{-1}(b\nu-d\mu),\\
\langle X\tilde{X}\rangle&=&\Delta^{-1}(ad-bc),\\
\langle \tilde{X}\tilde{X}\rangle&=&\Delta^{-1}(c\mu-a\nu),\\
\end{array}\]
where \[\Delta=b^2c^2-2abcd+a^2d^2+cd\mu^2-(bc+ad)\mu\nu+ab\nu^2.\] 
These match the A/2 correlation functions in 
equation~(\ref{A1}), if we identify $X$ with $\psi$ and
$\tilde{X}$ with $\tilde{\psi}$. 

We also checked that all 
four-point functions for general $A$, $B$, $C$, $D$
(as listed in appendix~\ref{app:p1p1}) match the results from the A/2 model. 
For the special case in which $\det B=\det C=0$, we have checked that all 
correlation funcions up to ten-point correlation functions and one 
twelve-point correlation function $\langle X^6\tilde{X}^6\rangle$ match the 
results from the A/2 model.

Beyond special cases, there is also a general argument that all correlation
functions must match.  We will utilize a formula for the A/2 model correlation
functions given in \cite{08100012}[section 3.4], which is similar in form
to the formula above for B/2 Landau-Ginzburg model correlation functions,
and argue that after some algebra,
the formula for A/2 correlation functions in
\cite{08100012} matches the formula for B/2 correlation functions above.
As a result,
all correlation functions in our B/2-twisted Landau-Ginzburg model
will necessarily match those of the A/2 nonlinear sigma model.

Let us describe this argument for general matching correlation functions.
From \cite{08100012}[section 3.4], all correlation functions in an
A/2-twisted (0,2) nonlinear sigma model on ${\mathbb P}^1 \times
{\mathbb P}^1$, at genus zero, take the form
\begin{eqnarray*}
\lefteqn{
\langle f(\psi, \tilde{\psi} ) \rangle
} \\
 & = &
\sum_{ \psi, \tilde{\psi} | {\cal J}_a=0 }
f(\psi, \tilde{\psi}) \left[
\left( \det_{a,b} {\cal J}_{a,b} \right)
\det \left( A \psi + B \tilde{\psi} \right)
\det \left( C \psi + D \tilde{\psi} \right) \right]^{-1} ,
\\
& = & 
\sum_{ \psi, \tilde{\psi} | {\cal J}_a=0 }
f(\psi, \tilde{\psi}) 
\det \left[
\begin{array}{cc}
\partial_{\psi} \det (A \psi + B \tilde{\psi}) &
\partial_{\tilde{\psi}} \det (A \psi + B \tilde{\psi}) \\
\partial_{\psi} \det (C \psi + D \tilde{\psi}) &
\partial_{\tilde{\psi}} \det ( C \psi + D \tilde{\psi})
\end{array} \right]^{-1}
\end{eqnarray*}
where the ${\cal J}_a$ (not to be confused with the $J$, $\tilde{J}$ we
used in our dual theory earlier) are defined by
\begin{eqnarray*}
{\cal J}_1 & = & \ln \left( q_1^{-1} \det 
\left( A \psi + B \tilde{\psi} \right)
\right), \\
{\cal J}_2 & = & \ln \left( q_2^{-1} \det 
\left( C \psi + D \tilde{\psi} \right)
\right).
\end{eqnarray*}

To compare the correlation functions above with the B/2 correlation functions
in our dual theory, which take a similar form, first note that the
constraint $\tilde{J}_a = 0$ implies
\begin{displaymath}
\det( A \psi + B \tilde{\psi}) \: = \: q_1, \: \: \:
\det( C \psi + D \tilde{\psi}) \: = \: q_2,
\end{displaymath}
the quantum sheaf cohomology relations and also the relations defining the
vacua of the B/2 Landau-Ginzburg model.
Then, matching follows as a consequence of
\begin{displaymath}
\det_{i,j} J_{i,j} \: = \: 
\left( \det_{a,b} {\cal J}_{a,b} \right)
\det \left( A \psi + B \tilde{\psi} \right)
\det \left( C \psi + D \tilde{\psi} \right),
\end{displaymath}
or more explicitly
\begin{eqnarray*}
\lefteqn{
\det\left[ \begin{array}{cc}
\partial_Y \left( a X + b \tilde{X}^2/X + \mu \tilde{X} - q_1/X\right) &
\partial_{\tilde{Y}} \left(
 a X + b \tilde{X}^2/X + \mu \tilde{X} - q_1/X\right) \\
\partial_Y \left( d \tilde{X} + c X^2/\tilde{X} + \nu X - q_2/\tilde{X} \right)
&
\partial_{\tilde{Y}} \left(
 d \tilde{X} + c X^2/\tilde{X} + \nu X - q_2/\tilde{X} \right)
\end{array} \right]
} \\
& \hspace*{1.5in}  = &
\det  \left[
\begin{array}{cc}
\partial_{\psi} \det (A \psi + B \tilde{\psi}) &
\partial_{\tilde{\psi}} \det (A \psi + B \tilde{\psi}) \\
\partial_{\psi} \det (C \psi + D \tilde{\psi}) &
\partial_{\tilde{\psi}} \det ( C \psi + D \tilde{\psi})
\end{array} \right],
\end{eqnarray*}
where $X = \exp(Y)$, $\tilde{X} = \exp(\tilde{Y})$,
after identifying $X$ with $\psi$ and $\tilde{X}$ with $\tilde{\psi}$,
which is straightforward to verify.
Thus, all genus zero correlation functions of our B/2 Landau-Ginzburg
model, the proposed dual to ${\mathbb P}^1 \times {\mathbb P}^1$,
do indeed match the correlation functions of the (0,2) theory
on ${\mathbb P}^1 \times {\mathbb P}^1$.

We will apply a more general version of this argument when checking
genus zero correlation functions of the proposed B/2 Landau-Ginzburg
dual to A/2 theories on ${\mathbb P}^n \times {\mathbb P}^n$ in
section~\ref{sect:pnpmdual}.

\subsection{Moduli}
\label{sect:p1p1-moduli}

On the face of it, the correlation functions above are determined by
six numbers:
\begin{displaymath}
\det A, \: \: \:
\det B, \: \: \:
\det C, \: \: \:
\det D, \: \: \:
\det(A+B), \: \: \:
\det(C+D),
\end{displaymath}
(in addition, of course, to $q_1$, $q_2$).
Not all of the individual elements of each of the four matrices
$A$, $B$, $C$, $D$ are pertinent, essentially because this theory admits
global $GL(2)$ actions rotating those matrices.  In addition, in principle
field redefinitions could be used to also eliminate some of the parameters
above.

Mathematically, the tangent bundle of ${\mathbb P}^1 \times {\mathbb P}^1$
also has six moduli (as counted in appendix~\ref{app:moduli}), 
matching the count above.  However, if one deforms 
to a finite distance away from the tangent bundle, the number of 
mathematical bundle moduli (counted by $H^1({\mathbb P}^1 \times {\mathbb P}^1,
{\rm End}\, {\cal E})$ for bundle ${\cal E}$)
may drop, as we discuss
in appendix~\ref{app:moduli}.  
Furthermore, not all of those moduli need necessarily be expressible
monadically, as polynomial deformations of the GLSM, so the true number
of parameters that the GLSM can access may be significantly smaller
(reflecting {\it e.g.} the symmetries and field redefinitions mentioned
above).  
We will see this in an example in section~\ref{sect:redundancy},
where we will take models with matrices $B$ such that
$\det B \neq 0$, and construct equivalent theories with
$\det B = 0$.

\subsection{Redundancies and equivalent descriptions}
\label{sect:redundancy}

As the moduli counts in the last section suggest, our description of the
theories in terms of four matrices $A$, $B$, $C$, $D$ has a great deal of
redundancy.  This can be expressed in the fact that there are
three $GL(2)$ actions\footnote{
We would like to thank R. Donagi for making this observation originally.
See also a related discussion in \cite{dlm}.} on these matrices.  Specifically,
three matrices
$P$, $Q$, $R$ each in 
$GL(2)$
act on the matrices $A$, $B$, $C$, $D$ as follows:
\begin{displaymath}
\left[ \begin{array}{cc}
A & B \\
C & D \end{array} \right] \: \mapsto \:
\left[ \begin{array}{cc}
P A & P B \\
Q C & Q D 
\end{array} \right] R
\end{displaymath}
at the same time that
\begin{displaymath}
\left[ \begin{array}{c} \psi \\ \tilde{\psi} \end{array} \right]
\: \mapsto \:
R \left[ \begin{array}{c} \psi \\ \tilde{\psi} \end{array} \right] ,
\end{displaymath}
and
\begin{displaymath}
q_1 \: \mapsto \: (\det P) q_1, \: \: \:
q_2 \: \mapsto \: (\det Q) q_2 .
\end{displaymath}
Of course, these three $GL(2)$ actions are not completely independent,
but in broad brushstrokes, they are the reason that there are no more
than six independent moduli yet sixteen naive parameters (the elements of
the four $2 \times 2$ matrices).

To understand how correlation functions behave, let us consider
a residue expression for correlation functions from \cite{1512.08058}:
\begin{eqnarray*}
\lefteqn{
\langle f(\psi, \tilde{\psi}) \rangle
} \\
& = &
\sum_{k_1, k_2} q_1^{k_1} q_2^{k_2} {\rm JKG-Res} \frac{
f(\psi, \tilde{\psi}) 
}{
( \det (A \psi + B \tilde{\psi}) )^{k_1+1} 
\det( C \psi + D \tilde{\psi}) )^{k_2+1}
} .
\end{eqnarray*}
Formally,
if we rotate $\psi$, $\tilde{\psi}$ by the matrix $R$ at the same time that
$A$, $B$, $C$, $D$ are also rotated by $R$, the new resulting expression
is equivalent to the original one, after a linear field redefinition.
In other words,
\begin{displaymath}
\langle f(R(\psi, \tilde{\psi})) \rangle_{R(A, B, C, D)} \: = \:
\frac{1}{| \det R | }
\langle f(\psi, \tilde{\psi}) \rangle_{A, B, C, D}.
\end{displaymath}
That said, the expressions for correlation functions we utilize in this
paper assume that $A$ and $D$ are both invertible, and a general $R$-rotation
could change that.  In such cases, the pole prescription implicit in the
definition of the JKG residue in \cite{1512.08058} 
would yield different results,
so one should be careful in applying the formal statement above.

An example of such equivalences is as follows.
Define $\beta$ to be a solution of
\begin{displaymath}
(\det A) \beta^2 + \mu \beta + (\det B) = 0,
\end{displaymath}
{\it i.e.}
\begin{displaymath}
\beta = \frac{1}{2a} \bigg( -\mu \pm \sqrt{\mu^2-4ab} \bigg) ,
\end{displaymath}
(where $A$ is assumed invertible,)
and let us assume $C=0$.
Take
\[
R=
\begin{bmatrix}
1 & \beta \\
0 & 1
\end{bmatrix}.
\]
This matrix $R$ rotates $B$ to a noninvertible matrix.  Specifically,
under the action of $R$, 
\begin{displaymath}
B \: \mapsto \: B' \: = \: B + \beta A,
\end{displaymath}
and the other matrices are invariant.
It is straightforward to check that $\det B' = 0$.
In principle, correlation functions in the original theory should match
correlation functions with these parameters so long as $\psi$, 
$\tilde{\psi}$ are suitably rotated:
\begin{displaymath}
\langle f(\psi, \tilde{\psi}) \rangle_{\rm original} \: = \:
\langle f(\psi + \beta \tilde{\psi}, \tilde{\psi} + \gamma \psi)
\rangle_{\rm new}.
\end{displaymath}

Now, having constructed an equivalent model for which
$\det B' = 0$,
we can construct the dual (0,2) Landau-Ginzburg theory.
This is defined by the (0,2) superpotential with
\begin{eqnarray*}
J' & = & a X +  \mu' \tilde{X} -
\frac{q_1}{X}, \\
\tilde{J}' & = & d \tilde{X} +  \nu' X -
\frac{q_2}{\tilde{X}},
\end{eqnarray*}
where
\begin{displaymath}
\mu' = \det(A+B') - \det A - \det B', \: \: \:
\nu' = \det(C+D) - \det C - \det D = \nu.
\end{displaymath}
This is just the specialization of our previous proposed dual to case with
$B'$ instead of $B$ and with $C=0$, so that the 
$\tilde{X}^2/X$ and $X^2/\tilde{X}$ terms vanish.

Given the rotation on the original $\psi$, $\tilde{\psi}$, we see that
in principle the original correlation functions should match the
correlation functions in the final Landau-Ginzburg model above as
\begin{displaymath}
\langle f(\psi, \tilde{\psi}) \rangle_{\rm original}
\: = \: 
\langle f(X + \beta \tilde{X}, \tilde{X} ) \rangle_{\rm final} 
.
\end{displaymath}

Now, let us turn to a particular special case, appearing
in \cite{0309226}.  This special case is the sole previous example
of a (0,2) Landau-Ginzburg mirror to a A/2-twisted theory that
had previously appeared in the literature.  More to the point, this sole
example in the literature does not fit the pattern we have discussed in
previous sections, and instead is related to them via a field redefinition
of the form discussed in this section.

Specifically, let us consider the case
\begin{displaymath}
A=D=I, \: \: \:
C=0, \: \: \:
B=\left[\begin{array}{cc}
\epsilon_1 & 0\\
0 & \epsilon_2\\
\end{array}\right].
\end{displaymath} 
Following the methods we have discussed prior to this section, the
dual Landau-Ginzburg theory 
has the parameters
\begin{displaymath}
a = 1, \: \: \:
b = \epsilon_1 \epsilon_2, \: \: \:
c = 0, \: \: \:
d = 1, \: \: \: 
\mu = \epsilon_1 + \epsilon_2, \: \: \:
\nu = 0,
\end{displaymath}
and superpotential
\begin{displaymath}
W \: = \: F \left( X + \epsilon_1  \epsilon_2 \frac{ \tilde{X}^2 }{X}
+ (\epsilon_1 + \epsilon_2) \tilde{X} - \frac{q_1}{X} \right)
\: + \:
\tilde{F} \left( \tilde{X} - \frac{q_2}{\tilde{X}} \right).
\end{displaymath}
The two-point correlation functions in this theory, for example, are
\begin{eqnarray*}
\langle X X \rangle & = & - (\epsilon_1 + \epsilon_2), \\
\langle X \tilde{X} \rangle & = & 1, \\
\langle \tilde{X} \tilde{X} \rangle & = & 0.
\end{eqnarray*}

Unfortunately, although this does correctly capture the A/2
correlation functions,
neither the superpotential nor the correlation functions above match
those given in \cite{0309226} as the dual.

To find the presentation of the dual given in \cite{0309226}, 
one must instead perform a $R$-rotation of the sort
described above, rotating $B$ and $C$ to noninvertible matrices. 
One then computes
\begin{eqnarray*}
\beta & = & \frac{1}{2} \left( - \epsilon_1 - \epsilon_2 +
(\epsilon_1 - \epsilon_2) \right) \: = \: - \epsilon_2, \\
\gamma & = & 0,
\end{eqnarray*}
(taking the positive square root in $\beta$).
After transforming by
\begin{displaymath}
R \: = \: \left[ \begin{array}{cc}
1 & \beta \\
\gamma & 1 \end{array} \right] \: = \:
\left[ \begin{array}{cc}
1 & - \epsilon_2 \\
0 & 1 \end{array} \right],
\end{displaymath}
one has the new dual defined by parameters
\begin{displaymath}
a' = 1, \: \: \:
b' = \det \left[ \begin{array}{cc}
\epsilon_1 - \epsilon_2 & 0 \\
0 & 0 \end{array} \right] = 0, \: \: \;
c' = 0, \: \: \:
d' = 1,
\end{displaymath}
\begin{displaymath}
\mu' = \det(A'+B') -  \det A' - \det B' = \epsilon_1 - \epsilon_2, 
\end{displaymath}
\begin{displaymath}
\nu' = \det(C'+D') - \det C' - \det D' = 0,
\end{displaymath}
hence the superpotential
\begin{displaymath}
W \: = \: F \left( X + (\epsilon_1 - \epsilon_2) \tilde{X} - 
\frac{ q_1 }{X} \right) \: + \:
\tilde{F} \left( \tilde{X} - \frac{ q_2 }{ \tilde{X} } \right).
\end{displaymath}
From the results in appendix~\ref{app:p1p1}, the two-point functions in
this Landau-Ginzburg model are given by
\begin{displaymath}
\langle X X \rangle \: = \: \epsilon_2 - \epsilon_1, \: \: \:
\langle X \tilde{X} \rangle \: = \: 1, \: \: \:
\langle \tilde{X} \tilde{X} \rangle \: = \: 0,
\end{displaymath}
matching the results of \cite{0309226}.

Also note that, in this same theory,
\begin{eqnarray*}
\langle (X - \epsilon_2 \tilde{X} )^2 \rangle & = &
\langle X^2 \rangle - 2 \epsilon_2 \langle X \tilde{X} \rangle
+ \epsilon_2^2 \langle \tilde{X}^2 \rangle \: = \:
- \epsilon_1 - \epsilon_2, 
\\
\langle (X - \epsilon_2 \tilde{X} ) \tilde{X} \rangle & = &
\langle X \tilde{X} \rangle - \epsilon_2 \langle \tilde{X}^2 \rangle
\: = \: 1, \\
\langle \tilde{X} \tilde{X} \rangle & = & 0,
\end{eqnarray*}
matching the correlation functions of the original A/2-twisted theory,
as expected.
In \cite{0710.2354}, the change of variables above was given to correlate
A/2 correlation functions with those of the proposed dual theory, and here
we see that this is a special case of a much more general redundancy in
the description.

\section{Generalization to $\mathbb{P}^n\times\mathbb{P}^m$}\label{pn}

\subsection{The A/2-twisted nonlinear sigma model}

Let us begin by briefly reviewing pertinent properties of the (0,2)
nonlinear sigma model on ${\mathbb P}^n \times {\mathbb P}^m$, whose
dual we shall describe.  First, the gauge bundle in this (0,2) theory is
a deformation
${\cal E}$ of the tangent bundle of 
$\mathbb{P}^n\times\mathbb{P}^m$, which can be described as a cokernel
\begin{displaymath}
0 \: \longrightarrow \: {\cal O}^2 \: \stackrel{E}{\longrightarrow} \:
{\cal O}(1,0)^{n+1} \oplus {\cal O}(0,1)^{m+1} \: \longrightarrow \: {\cal E}
\: \longrightarrow \: 0,
\end{displaymath}
where
\begin{displaymath}
E \: = \: \left[ \begin{array}{cc}
A & B \\
C & D
\end{array} \right],
\end{displaymath} 
in which $A$, $B$ are $(n+1)\times (n+1)$ matrices and
$C$, $D$ are $(m+1) \times (m+1)$ matrices.
The quantum sheaf cohomology ring of an A/2-twisted nonlinear sigma
model on ${\mathbb P}^n \times {\mathbb P}^m$ with the bundle above
takes the form \cite{08100012,1110.3751,1110.3752,1512.08058}
\begin{displaymath}
\det (A \psi + B \tilde{\psi}) \: = \: q_1, \: \: \:
\det (C \psi + D \tilde{\psi}) \: = \: q_2,
\end{displaymath}
and for later use, we expand the determinants as follows:
\begin{align}
\det(A\psi + B\tilde{\psi})
&=a\psi^{n+1}+b\tilde{\psi}^{n+1}+
\sum_{i=1}^{n}\mu_i \psi^i\tilde{\psi}^{n+1-i},\\
\det(C\psi+D\tilde{\psi})
&=c \psi^{m+1} + d\tilde{\psi}^{m+1}+
\sum_{k=1}^{m}\nu_k \psi^k\tilde{\psi}^{m+1-k},
\end{align}
where 
\begin{displaymath}
a = \det A, \: \: \:
b = \det B, \: \: \:
c = \det C, \: \: \:
d = \det D, \: \: \:
\end{displaymath}
$\mu_i$ is a sum of determinants of matrices, each of which is 
formed by taking $i$ rows of $A$ and $n+1-i$ rows of $B$, and
$\nu_i$ is formed similarly from $C$, $D$.

\subsection{The Toda-like mirror theory}
\label{sect:pnpmdual}

We claim the (0,2) superpotential of the (0,2) Landau-Ginzburg 
Toda-like mirror to $\mathbb{P}^n\times\mathbb{P}^m$ is
\begin{equation}
W=\sum_{i=1}^{n}F_iJ_i+\sum_{k=1}^{m}\tilde{F}_k\tilde{J}_k,
\end{equation}
where \begin{align}
J_i & =a^{(1-n)/n}\left(aX_i-\frac{q_1}{X_1\cdots X_n}+b\frac{\tilde{X}_1^{n+1}}{X_1^n}+\sum_{i=1}^{n}\mu_{n+1-i}\frac{\tilde{X}_1^i}{X_1^{i-1}}\right),\\
\tilde{J}_k & =d^{(1-m)/m}\left(d\tilde{X}_k-\frac{q_2}{\tilde{X}_1
\cdots \tilde{X}_m}+c\frac{X_1^{m+1}}{\tilde{X}_1^m} +
\sum_{k=1}^{m}\nu_k\frac{X_1^k}{\tilde{X}_1^{k-1}}\right),
\end{align} 
which clearly generalizes the dual to
${\mathbb P}^1 \times {\mathbb P}^1$ discussed in
section~\ref{sect:p1p1lg}.

First, note that if the parameters $a$, $b$, $c$, $d$, and the $\mu_i$, $\nu_k$
are related to the matrices $A$, $B$, $C$, $D$ of the A/2 model as above,
then the vacua of this theory, defined by
$J_i = 0 = \tilde{J}_k$, are the solutions of
\begin{displaymath}
X_1 \: = \: X_2 \: = \: \cdots \: = \: X_n \: \equiv \: X, \: \: \:
\tilde{X}_1 \: = \: \tilde{X}_2 \: = \: \cdots \: = \: \tilde{X}_m 
\: \equiv \: \tilde{X},
\end{displaymath}
\begin{displaymath}
\det(A X + B \tilde{X}) \: = \: q_1, \: \: \:
\det(C X + D \tilde{X}) \: = \: q_2,
\end{displaymath}
identical to 
the solutions of the quantum sheaf cohomology relations, as one would expect
for a sensible Toda-like dual.

One can show that the correlation functions of this B/2-twisted 
Landau-Ginzburg model computed by equation~(\ref{B/2 corr.}) 
equal the correlation functions of A/2-twisted model on 
$\mathbb{P}^n\times\mathbb{P}^m$ \cite{08100012}:
\begin{equation}
\langle \sigma_{a_1}\cdots\sigma_{a_l}\rangle
=
\sum_{\sigma|{\cal J}=0}^{}\sigma_{a_1}\cdots\sigma_{a_l}
\left[\det\limits_{a,b}{\cal J}_{a,b}\prod_{\alpha}^{}\det M_{(\alpha)}
\right]^{-1}
\end{equation}
with \begin{equation}
{\cal J}_a=
\ln\left[q_a^{-1}\prod\limits_{\alpha}^{}\det M_{(\alpha)}^{Q^a_{(\alpha)}}\right].
\end{equation}
In the present case, for ${\mathbb P}^n \times {\mathbb P}^m$,
there are only two $\sigma$'s, which we label $\sigma_1$, $\sigma_2$,
and
\begin{eqnarray*}
{\cal J}_1 & = & \ln \left[ q_1^{-1} \det( A \sigma_1 + B \sigma_2 ) \right] ,\\
{\cal J}_2 & = & \ln \left[ q_2^{-1} \det( C \sigma_1 + D \sigma_2 ) \right].
\end{eqnarray*}

To show that the two expressions for correlation functions
match, it suffices to show that 
\begin{equation}  \label{eq:pnxpm-pf}
\det|J_{i,j}|=
\det\limits_{a,b}|{\cal J}_{a,b}|\prod_{\alpha}^{}\det M_{(\alpha)}, 
\end{equation}
by identifying $X_i$ with $\sigma_{1}$ and $\tilde{X}_k$ 
with $\sigma_{2}$ on the space of vacua, since 
\begin{align} \label{eq:pnxpm-sigma}
\det\limits_{a,b}|{\cal J}_{a,b}|\prod_{\alpha}^{}\det M_{(\alpha)}=\det
\begin{bmatrix}
\partial_{\sigma_1}\det(A\sigma_1+B\sigma_2) & \partial_{\sigma_2}\det(A\sigma_1+B\sigma_2)\\
\partial_{\sigma_1}\det(C\sigma_1+D\sigma_2) & \partial_{\sigma_2}\det(C\sigma_1+D\sigma_2)\\
\end{bmatrix}
\end{align}
on the classical vacua ${\cal J}_a(\sigma)=0$.

In order to show~(\ref{eq:pnxpm-pf}), we will need a minor linear algebra
result.
For an $(n+m)\times (n+m)$ matrix of the form
\begin{equation} \label{eq:pnxpm-matrix}
\begin{matrix} 
\coolleftbrace{n}{a \\ a\\ a \\ a \\ a}\\
\coolleftbrace{m}{c \\ 0 \\ 0 \\ 0 \\0}
\end{matrix}
\begin{bmatrix}
a_{11} & a_{12} & a_{13} & \cdots & a_{1n} & \beta & 0 & & \cdots & 0\\
-\alpha & \alpha & 0 & \cdots & 0 & 0 & 0 & & \cdots & 0\\
-\alpha & 0 & \alpha & \cdots & 0 & 0 & & & & 0\\
\vdots & & & \ddots & \vdots & \vdots & & & \ddots & \vdots\\
-\alpha & 0 & \cdots & 0 & \alpha & 0 & & & \cdots & 0\\
\rho & 0 & & \cdots & 0 & d_{11} & d_{12} & d_{13} & \cdots & d_{1m}\\
0 & 0 & & \cdots & 0 & -\delta & \delta & 0 & \cdots & 0\\
0 & & & & 0 & -\delta & 0 & \delta & \cdots & 0\\
\vdots & & & \ddots & \vdots & \vdots & & & \ddots & \vdots\\
0 & & & \cdots & 0 & -\delta & 0 & \cdots & 0 & \delta
\end{bmatrix},
\end{equation}
its determinant has the form
\begin{equation}
(\det \zeta) (\det \eta) - \beta \rho \alpha^{n-1} \delta^{m-1} ,
\end{equation}
where $\det \zeta$ is the determinant of the upper-left $n\times n$ submatrix 
and $\det \eta$ is the determinant of the lower-right $m \times m$ submatrix,
given by 
\begin{align}
\det \zeta&=
\alpha^{n-1}
\sum\limits_{i=1}^{n}a_{1i} ,\\
\det \eta&=
\delta^{m-1}
\sum\limits_{k=1}^{m}d_{1k} .
\end{align}

Next, we need to compute
\begin{displaymath}
\det|J_{i,j}|
=\det
\begin{bmatrix}
\partial_{Y_1}J_1 & \cdots & \partial_{Y_n}J_1 & \partial_{\tilde{Y}_1}J_1 & 
\cdots & \partial_{\tilde{Y}_m}J_1\\
\vdots & \ddots & \vdots & \vdots & \ddots &\vdots\\
\partial_{Y_1}J_n & \cdots & \partial_{Y_n}J_n & \partial_{\tilde{Y}_1}J_n & 
\cdots & \partial_{\tilde{Y}_m}J_n\\
\partial_{Y_1}\tilde{J}_1 & \cdots & \partial_{Y_n}\tilde{J}_1 & 
\partial_{\tilde{Y}_1}\tilde{J}_1 & \cdots & \partial_{\tilde{Y}_m}\tilde{J}_1\\
\vdots & \ddots & \vdots & \vdots & \ddots &\vdots\\
\partial_{Y_1}\tilde{J}_m & \cdots & \partial_{Y_n}\tilde{J}_m & 
\partial_{\tilde{Y}_1}\tilde{J}_m & \cdots & \partial_{\tilde{Y}_m}\tilde{J}_m\\
\end{bmatrix},
\end{displaymath}
where $X_i = \exp(Y_i)$ and $\tilde{X}_i = \exp(\tilde{Y}_i)$.
By taking suitable linear combinations, one can rewrite the matrix above
in the form of the matrix~(\ref{eq:pnxpm-matrix}),
with the following identifications:
\begin{align*}
a_{11} &= a^{(1-n)/n}\left(
aX_1+\frac{q_1}{X_1\cdots X_n}-nb\frac{\tilde{X}_1^{n+1}}{X_1^n}+
\sum_{i=1}^{n}(1-i)\mu_{n+1-i}\frac{\tilde{X}_1^i}{X_1^{i-1}}\right), \\
&= a^{(1-n)/n}\left(2aX + (1-n)b\frac{\tilde{X}^{n+1}}{X^n}+
\sum_{i=1}^n(2-i)\mu_{n+1-i}\frac{\tilde{X}^i}{X^{i-1}}\right), \\ 
a_{12} &= a_{13} = \cdots = a_{1n}= a^{(1-n)/n}\left(+\frac{q_1}{X_1\cdots X_n}
\right), \\
&= a^{(1-n)/n}\left(
aX + b\frac{\tilde{X}^{n+1}}{X^n}+
\sum_{i=1}^n \mu_{n+1-i} \frac{\tilde{X}^i}{X^{i-1}}\right),   
\end{align*}
\begin{align*}
\alpha &= a^{(1-n)/n}(aX_1), \\
&= a^{1/n}X, \\
\beta &=a^{(1-n)/n}\left(
(n+1)b\frac{\tilde{X}_1^{n+1}}{X_1^n} + 
\sum_{i=1}^n i \mu_{n+1-i}\frac{\tilde{X_1^i}}{X_1^{i-1}}\right), \\
&= a^{(1-n)/n}\left(
(n+1)b\frac{\tilde{X}^{n+1}}{X^n}+
\sum_{i=1}^n i \mu_{n+1-i} \frac{\tilde{X}^i}{X^{i-1}}\right),
\end{align*}
\begin{align*}
d_{11} &= d^{(1-m)/m}\left(
d\tilde{X}_1 + \frac{q_2}{\tilde{X}_1 \cdots \tilde{X}_m}
-c m \frac{X_1^{m+1}}{\tilde{X}_1^m} + 
\sum_{k=1}^m (1-k)\nu_k \frac{X_1^k}{\tilde{X}_1^{k-1}}\right), \\
&= d^{(1-m)/m} \left(
2d\tilde{X} + (1-m)c\frac{X^{m+1}}{\tilde{X}^m} + 
\sum_{k=1}^m (2-k)\nu_k \frac{X^k}{\tilde{X}^{k-1}}\right), \\
d_{12} &= d_{13} = \cdots = d_{1n}= d^{(1-m)/m}\left(
+\frac{q_2}{\tilde{X}_1\cdots \tilde{X}_m}\right), \\
&= d^{(1-m)/m}\left(
d\tilde{X} + c\frac{X^{m+1}}{\tilde{X}^m} + 
\sum_{k=1}^m \nu_k \frac{X^k}{\tilde{X}^{k-1}}\right), 
\end{align*}
\begin{align*}
\delta &= d^{(1-m)/m}(d\tilde{X}_1),\\
&= d^{1/m}\tilde{X}, \\
\rho &=d^{(1-m)/m}\left(
(m+1)c\frac{X_1^{m+1}}{\tilde{X}_1^m} + 
\sum_{k=1}^m k \nu_{k} \frac{X_1^k}{\tilde{X}_1^{k-1}} \right),\\
&= d^{(1-m)/m}\left(
(m+1)\frac{X^{m+1}}{\tilde{X}^m}+
\sum_{k=1}^m k \nu_k \frac{X^k}{\tilde{X}^{k-1}}\right).
\end{align*}
(In the expressions above, the second line is obtained by evaluation
on vacua.)

Putting this together, we can write
\begin{eqnarray*}
\det |J_{i,j}| & = &
\det  \begin{bmatrix}
  \det \zeta & \beta \alpha^{n-1}\\
  \rho \delta^{m-1} & \det \eta 
 \end{bmatrix} \\
& = & 
\det \begin{bmatrix}
\alpha^{n-1} (a_{11} + (n-1) a_{12}) & 
\alpha^{n-1} \beta \\
\delta^{m-1} \rho &
\delta^{m-1} (d_{11} + (n-1) d_{12} )
\end{bmatrix}
\end{eqnarray*}
which is easily checked to be the determinant of
\begin{displaymath}
\left[
\begin{array}{l}
 (n+1)aX^n + \sum_{i=1}^n (n+1-i) \mu_{n+1-i} \tilde{X}^i X^{n-i} \\
\hspace*{1.8in}
(n+1) b \tilde{X}^{n+1} X^{-1} + \sum_{i=1}^n i \mu_{n+1-i} \tilde{X}^i X^{n-i} \\
 (m-1)cX^{m+1} \tilde{X}^{-1} + \sum_{k=1}^m k \nu_k X^k \tilde{X}^{m-k} \\
\hspace*{1.8in}
(m-1)d\tilde{X}^m + \sum_{k=1}^m (m+1-k) \nu_k X^k \tilde{X}^{m-k}
\end{array}
\right].
\end{displaymath}
By identifying $X_i$ with $\sigma_{1}$ and
$\tilde{X}_k$
with $\sigma_{2}$,
we see that the determinant above matches
(\ref{eq:pnxpm-sigma}).

Thus, all genus-zero correlation functions in our proposed Toda dual
match those of the (0,2) theory on ${\mathbb P}^n \times {\mathbb P}^m$
with a deformation of the tangent bundle.
In addition to constructing a general argument that correlation functions
should match, we have also compared correlation functions in special cases,
as we shall outline next.

\subsection{Example:   $\mathbb{P}^1\times\mathbb{P}^2$}

As a consistency check, as we have already studied the dual to
${\mathbb P}^1 \times {\mathbb P}^1$,
we next consider the special case
${\mathbb P}^1 \times {\mathbb P}^2$.
Specializing the results for ${\mathbb P}^n \times {\mathbb P}^m$,
the mirror (0,2) Landau-Ginzburg model is defined by the superpotential
\begin{align*}
W=FJ+\widetilde{F_1}\widetilde{J_1}+\widetilde{F_2}\widetilde{J_2}
\end{align*}
with
\begin{align}
J&= aX- \frac{q_1}{X} + b\frac{\tilde{X}_1^2}{X} + \mu\tilde{X}_1,\\
\widetilde{J_1}&=d^{-\frac{1}{2}}\left( d\tilde{X}_1-
\frac{q_2}{\tilde{X}_1\tilde{X}_2} + c\frac{X^3}{\tilde{X}_1^2}
+ fX + g\frac{X^2}{\tilde{X}_1} \right),\\
\widetilde{J_2}&=d^{-\frac{1}{2}}\left( d\tilde{X}_2 - 
\frac{q_2}{\tilde{X}_1
\tilde{X}_2} + c\frac{X^3}{\tilde{X}_1^2} + fX + g\frac{X^2}{\tilde{X}_1}
\right) .
\end{align}
In the expression above,
\[\begin{array}{rcl}
a&=&\det A, \quad b=\det B, \quad c=\det C, \quad d=\det D,\\
\mu&=&\det(A+B)-\det A-\det B ,
\end{array}\]
for the matrices $A$, $B$, $C$, $D$ defining the gauge bundle
deformation in the A/2-twisted nonlinear sigma model, and where
$g$ is a sum of determinants of three matrices, each of which is 
formed from taking two rows of $C$ and one row of $D$, 
and $f$ is similarly a sum of three determinants, involving
matrices 
formed as two rows of $D$ and one row of $C$.

We have directly computed correlation functions in the proposed dual
Landau-Ginzburg theory above, in the special case $c=f=g=0$.
On the vacua, $\tilde{X}_1 = \tilde{X}_2$, so in computing correlation
functions, we will use $\tilde{X}$ to denote either
$\tilde{X}_1$ or $\tilde{X}_2$.
In any event, the three-point correlation functions in this case are
given by
\[\begin{array}{rcl}
\langle XXX\rangle&=& -(ab-\mu^2)(a^3d)^{-1},\\
\langle XX\tilde{X}\rangle&=& -\mu(a^2d)^{-1},\\
\langle X\tilde{X} \tilde{X}\rangle&=& (ad)^{-1},\\
\langle \tilde{X} \tilde{X} \tilde{X} \rangle&=& 0.
\end{array}
\]

The five-point correlation functions are given by
\[\begin{array}{rcl}
\langle X^5\rangle&=& -(a^4d)^{-1}(2ab-3\mu^2)q_1,\\
\langle X^4\tilde{X}\rangle&=& -2(a^3d)^{-1}\mu q_1,\\
\langle X^3\tilde{X}^2\rangle&=& (a^2d)^{-1}q_1,\\
\langle X^2\tilde{X}^3\rangle&=& 0,\\
\langle X\tilde{X}^4\rangle&=& 0,\\
\langle \tilde{X}^5\rangle&=& 0.
\end{array}
\]

The six-point correlation functions are given by
\[\begin{array}{rcl}
\langle X^6\rangle&=& -(a^6d^2)^{-1}\mu(3a^2b^2-4ab\mu^2+\mu^4)q_2,\\
\langle X^5\tilde{X}\rangle&=& (a^5d^2)^{-1}(a^2b^2-3ab\mu^2+\mu^4)q_2,\\
\langle X^4\tilde{X}^2\rangle&=& -(a^4d^2)^{-1}\mu(-2ab+\mu^2)q_2,\\
\langle X^3\tilde{X}^3\rangle&=& -(a^3d^2)^{-1}(ab-\mu^2)q_2,\\
\langle X^2\tilde{X}^4\rangle&=& -(a^2d^2)^{-1}\mu q_2,\\
\langle X\tilde{X}^5\rangle&=& (ad^2)^{-1}q_2,\\
\langle \tilde{X}^6\rangle&=& 0.
\end{array}
\]

If we identify $X$ with $\psi$ and $\tilde{X}$ with
$\tilde{\psi}$, then these correlation functions match
those of the corresponding A/2 model, for this case ($c=f=g=0$).
We have listed the A/2 model correlation functions (for the general case)
in appendix~\ref{app:p1p2}.

\section{Conclusions}

In this paper we establish the (0,2) Toda-like dual models to
(0,2) nonlinear sigma models on 
$\mathbb{P}^n\times\mathbb{P}^m$ with a deformation of the tangent bundle,
solving an old problem on the road to understanding (0,2) mirror symmetry. 
We checked our ansatz via a general argument demonstrating that
all genus zero correlation functions match, and also checked
matching of low-order correlation functions explicitly.

We have only checked our ansatz for duals at genus zero.  It would
be useful to check at higher genera, but unfortunately at this time it
is not known how to compute higher genus correlation functions in A/2
twisted theories, so such checks are left for the future.

The methods used here, such as our use of quantum sheaf cohomology to
determine the vacua of the correct dual theory, reminiscent of
methods in {\it e.g.} \cite{iritani1},
should be straightforward to extend to more general
Fano toric varieties.

\section{Acknowledgements}

We would like to thank L.~Anderson, R.~Donagi, J.~Gray, J.~Guffin,
S.~Katz, Z.~Lu, and I.~Melnikov for useful conversations.
E.S. was partially supported by NSF grant PHY-1417410.

\appendix

\section{Correlation functions in some examples}

\subsection{A/2 correlation functions on 
${\mathbb P}^1 \times {\mathbb P}^1$}
\label{app:p1p1-a2-results}

In this appendix we list the two- and four-point correlation functions
for A/2 twisted nonlinear sigma models on ${\mathbb P}^1 \times
{\mathbb P}^1$ with a deformation ${\cal E}$ of the
tangent bundle, defined by
\[
0 \: \longrightarrow \: {\cal O}\otimes{\cal O} \: \xrightarrow{E} \: 
{\cal O}(1,0)^2\oplus{\cal O}(0,1)^2 \: \longrightarrow \: 
{\cal E} \: \longrightarrow \: 0 ,
\]
defined as in section~\ref{sect:p1p1-a2} by four matrices $A$, $B$, $C$, $D$.

In writing the correlation functions, we use the following notation:
\begin{displaymath}
a=\det(A), b=\det(B), c=\det(C), d=\det(D),
\end{displaymath}
\begin{displaymath}
e=\det(A+B),\quad f=\det(C+D),
\end{displaymath}
\begin{displaymath}
\mu = e - a - b, \: \: \:
\nu = f - c - d,
\end{displaymath}
\begin{eqnarray*}
\phi_1&=& \nu b - \mu d \: = \: ad+bf-de-bc,\\
\phi_2&=& ad-bc,\\
\phi_3&=& \mu c - \nu a \: = \: ad+ce-af-bc,
\end{eqnarray*}
\begin{eqnarray*}
\Delta & = & \phi_2^2 - \phi_1 \phi_3, \\
& =& (c-d)(bc-ad)e+cde^2+(a-b)(ad-bc)f-(bc+ad)ef+abf^2.
\end{eqnarray*}

The two-point correlation functions are given by
\begin{equation}
\langle \psi\psi\rangle = \frac{\phi_1}{\Delta}, \: \: \: 
\langle \psi\tilde{\psi}\rangle = \frac{\phi_2}{\Delta}, \: \: \: 
\langle \tilde{\psi}\tilde{\psi}\rangle = \frac{\phi_3}{\Delta}  .
\end{equation}
The four-point correlation functions are given by
\begin{align*}
\langle \psi\psi\psi\psi\rangle_{10}&=
 \frac{\phi_1}{\Delta^2} \left( \nu \phi_1 + 2 \phi_2 d \right)
= \frac{1}{\Delta^2}(\phi_1((f-c)\phi_1+ad^2+d^2e-bcd-bdf)),\\
\langle \psi\psi\psi\tilde{\psi}\rangle_{10}&=
\frac{1}{\Delta^2}\left(- \phi_1^2 c + \phi_2^2 d \right), \\
\langle \psi\psi\tilde{\psi}\tilde{\psi}\rangle_{10}&=
\frac{\phi_2}{\Delta^2} \left( \phi_3 d - \phi_1 c \right), \\
\langle \psi\tilde{\psi}\tilde{\psi}\tilde{\psi}\rangle_{10}&=
\frac{1}{\Delta^2} \left( \phi_3^2 d - \phi_2^2 c \right), \\
\langle \tilde{\psi}\tilde{\psi}\tilde{\psi}\tilde{\psi}\rangle_{10}&=
\frac{\phi_3}{\Delta^2} \left( \nu \phi_3 + 2 \phi_2 c \right), \\
&=\frac{1}{\Delta^2}(\phi_3(ce(c+d-f)+bc(c-d+f)+a((d-f)^2-c(d+f)))),\\
\langle \psi\psi\psi\psi\rangle_{01}&=
- \frac{\phi_1}{\Delta^2}\left(  \mu \phi_1 + 2 \phi_2 b \right), \\
&=\frac{1}{\Delta^2}(-\phi_1(2b(ad-bc)-d(a
+b-e)^2+b(a+b-e)(c+d-f))),\\
\langle \psi\psi\psi\tilde{\psi}\rangle_{01}&=
\frac{1}{\Delta^2}\left( \phi_1^2 a - \phi_2^2 b \right), \\
\langle \psi\psi\tilde{\psi}\tilde{\psi}\rangle_{01}&=
\frac{\phi_2}{\Delta^2}\left( - \phi_3 b + \phi_1 a \right),\\
\langle \psi\tilde{\psi}\tilde{\psi}\tilde{\psi}\rangle_{01}&=
\frac{1}{\Delta^2}\left( - \phi_3^2 b + \phi_2^2 a \right), \\
\langle \tilde{\psi}\tilde{\psi}\tilde{\psi}\tilde{\psi}\rangle_{01}&=
- \frac{\phi_3}{\Delta^2} \left(  \mu \phi_3  - 2 \phi_2 a \right)
=\frac{1}{\Delta^2}(\phi_3((e-b)\phi_3+a^2d+a^2f-abc-acd)),
\end{align*}
where the subscripts 10 and 01 denote contributions from the degree one sector
on either ${\mathbb P}^1$ factor:\[
\langle {\cal O}_1{\cal O}_2{\cal O}_3{\cal O}_4\rangle=q_1\langle {\cal O}_1{\cal O}_2{\cal O}_3{\cal O}_4\rangle_{10}+q_2\langle {\cal O}_1{\cal O}_2{\cal O}_3{\cal O}_4\rangle_{01}.\]

\subsection{Toda-like dual to
${\mathbb P}^1 \times {\mathbb P}^1$}
\label{app:p1p1}

In this appendix we list the two-point and four-point
correlation functions for our proposed Toda-like dual (0,2) Landau-Ginzburg
model, with superpotential of the form
\begin{align*}
J&=aX+b\frac{\tilde{X}^2}{X}+\mu\tilde{X}-\frac{q_1}{X},\\
\tilde{J}&=d\tilde{X}+c\frac{X^2}{\tilde{X}}+\nu X-\frac{q_2}{\tilde{X}} .
\end{align*}

The two-point correlation functions
in this (0,2) Landau-Ginzburg model can be shown to be
\[\begin{array}{rcl}
\langle XX\rangle&=&\gamma^{-1}(b\nu-d\mu),\\
\langle X \tilde{X}\rangle&=&\gamma^{-1}(ad-bc),\\
\langle \tilde{X} \tilde{X}\rangle&=&\gamma^{-1}(c\mu-a\nu),\\
\end{array}\]
where $\gamma=b^2c^2-2abcd+a^2d^2+cd\mu^2-(bc+ad)\mu \nu+ab\nu^2$.

The four-point correlation functions
in this (0,2) Landau-Ginzburg model can be shown to be
\[\begin{array}{rcl}
\langle XXXX\rangle_{10}&=&\gamma^{-2}(-(d\mu-b\nu)(d(2ad-\mu \nu)+b(-2cd+\nu^2))),\\
\langle XXXX\rangle_{01}&=&\gamma^{-2}(-d\mu+b\nu)(2b^2c+d\mu^2-b(2ad+\mu \nu)),\\
\langle XXX\tilde{X}\rangle_{10}&=&\gamma^{-2}(d((bc-ad)^2-cd\mu^2)+2bcd\mu \nu-b^2c\nu^2),\\
\langle XXX\tilde{X}\rangle_{01}&=&\gamma^{-2}(-b^3c^2+ad^2\mu^2-adb(ad+2\mu \nu)+ab^2(2cd+\nu^2)),\\
\langle XX\tilde{X} \tilde{X} \rangle_{10}&=&\gamma^{-2}(bc-ad)(-2cd\mu+bc\nu+ad\nu),\\
\langle XX\tilde{X} \tilde{X}\rangle_{01}&=&\gamma^{-2}(bc-ad)(bc\mu+ad\mu-2ab\nu),\\
\langle X\tilde{X} \tilde{X} \tilde{X}\rangle_{10}&=&\gamma^{-2}(c(-(bc-ad)^2+cd\mu^2)-2acd\mu \nu+a^2d\nu^2),\\
\langle X \tilde{X} \tilde{X} \tilde{X}\rangle_{01}&=&\gamma^{-2}(a^3d^2-bc^2\mu^2+abc(bc+2\mu \nu)-a^2b(2cd+\nu^2)),\\
\langle \tilde{X} \tilde{X} \tilde{X} \tilde{X}\rangle_{10}&=&\gamma^{-2}(c\mu-a\nu)(2bc^2-c\mu \nu+a(-2cd+\nu^2)),\\
\langle \tilde{X} \tilde{X} \tilde{X} \tilde{X}\rangle_{01}&=&\gamma^{-2}(c\mu-a\nu)(2a^2d+c\mu^2-a(2bc+\mu \nu)),\\
\end{array}
\]
where the $10$ and $01$ subscripts indicate the coefficients of
$q_1$, $q_2$, as in the previous subsection.

As remarked in section~\ref{sect:p1p1lg}, if we identify the parameters above
with matrix determinants as
\[\begin{array}{rcl}
a&=&\det A, \quad b=\det B, \quad c=\det C, \quad d=\det D,\\
\mu&=&\det(A+B)-\det A-\det B,\\
\nu&=&\det(C+D)-\det C-\det D,\\
\end{array}\]
for $A$, $B$, $C$, $D$ the matrices appearing in the A/2-twisted (0,2)
model on ${\mathbb P}^1 \times {\mathbb P}^1$, the correlation functions
in the Landau-Ginzburg model above match those of the A/2 model.

\subsection{A/2 correlation functions on
$\mathbb{P}^1\times\mathbb{P}^2$}
\label{app:p1p2}

In this appendix we list the three-point, five-point and six-point correlation 
functions for A/2 twisted nonlinear sigma models on 
${\mathbb P}^1 \times {\mathbb P}^2$ with a deformation 
${\cal E}$ of the tangent bundle,
defined by
\begin{displaymath}
0 \: \longrightarrow \: {\cal O}^2 \: \stackrel{E}{\longrightarrow} \:
{\cal O}(1,0)^2 \oplus {\cal O}(0,1)^3 \: \longrightarrow \:
{\cal E} \: \longrightarrow \: 0,
\end{displaymath}
with $A$, $B$ $2 \times 2$ matrices and $C$, $D$ $3 \times 3$ matrices.

Correlation functions in this theory can be computed in a variety of
methods, such as {\it e.g.} residues \cite{1512.08058}.
In writing the correlation functions, we use the following notation:
\begin{displaymath}
a  =  \det A, \: \: \:
b = \det B, \: \: \:
c = \det C, \: \: \:
d = \det D,
\end{displaymath}
\begin{displaymath}
\mu = \det(A+B) - \det A - \det B,
\end{displaymath}
$g$ is a sum of determinants of three matrices, each formed from two
rows of $C$ and one row of $D$, and $f$ is similarly a sum of three
determinants, each having two rows of $D$ and one row of $C$.

Three-point functions in the A/2 theory are given by
\begin{eqnarray*}
\langle \psi^3 \rangle &=& \Delta^{-1}(-a b d + b^2 g - b f \mu + d \mu^2), \\
\langle \psi^2 \tilde{\psi} \rangle &= & \Delta^{-1}
(-b^2 c + a b f - a d \mu), \\
\langle \psi \tilde{\psi}^2 \rangle &= & \Delta^{-1}
(a^2 d - a b g + b c \mu), \\
\langle \tilde{\psi}^3 \rangle &= & \Delta^{-1}
(a b c - a^2 f + a g \mu - c \mu^2),
\end{eqnarray*}
where
\begin{eqnarray*}
\lefteqn{
\Delta =a^3 d^2 + b \left((b c - a f)^2 - 2 a^2 d g + a b g^2 \right) 
 + (b c f + a d g) \mu^2 - c d \mu^3
}
\\
& & \hspace*{2in}
- (a d (-3 b c + a f) + b (b c + a f) g) \mu
.
\end{eqnarray*}

Five-point correlation functions in the A/2 theory are given by
\begin{eqnarray*}
\langle \psi^5 \rangle & =& q_1 \Delta^{-2} 
\left(b^4 (c^2 d - 2 c f g + g^3)
+ d^2 \mu^2 (3 a^2 d - 2 a f \mu + g \mu^2) \right. \\
& & + 2 b^3 (a g (f^2 - 2 d g) + (c f^2 + c d g - f g^2) \mu) \\
& &  +  b^2 (a^2 d (-f^2 + 5 d g) - 2 a f (f^2 - d g) \mu 
 + (-4 c d f + g (f^2 + 2 d g)) \mu^2)\\
& &\left.  -  2 b d (a^3 d^2 + a^2 d f \mu - 2 a (f^2 - d g) \mu^2
 + (-c d + f g)\mu^3)\right), \\
\\
\langle \psi^4 \tilde{\psi} \rangle & =& q_1 \Delta^{-2} 
\left(2 a^3 d^2 (b f - d \mu) + 2 a b^2 c (-b f^2 + 2 b d g + d f \mu) \right.\\
& &  +  a^2 (b^2 (-3 c d^2  
 + f^3 - 2 d f g) + 2 b d (-f^2 + d g) \mu + d^2 f \mu^2)\\
& &  +  c (b^4 (c f - g^2)
  - b^2 (f^2 + 2 d g) \mu^2 
 + 2 b d f \mu^3 - d^2 \mu^4\\
& &\left. + b^3 (-2 c d \mu + 2 f g \mu))\right), \\
\\
\langle \psi^3 \tilde{\psi}^2 \rangle & =& q_1 \Delta^{-2} \left(a^4 d^3
 - 2 a^3 b d^2 g + b^2 c^2 (b^2 g - 2 b f \mu + 3 d \mu^2)
\right. \\
& & +  a^2 (b^2 (2 c d f - f^2 g 
 + d g^2) + 2 b d f g \mu - d^2 g \mu^2)\\
& & \left.  - 2 a c (b^3 c d + 2 b d f \mu^2
 - d^2 \mu^3 + b^2 (-f^2 \mu + d g \mu))\right),\\
\\
\langle \psi^2 \tilde{\psi}^3 \rangle & =& q_1 \Delta^{-2} \left(-b^4 c^3 
+ 2 a b^3 c^2 f - a^2 d^2 (a^2 f - 2 a g \mu + 3 c \mu^2) \right.\\
& & +  b^2 (a^2 (f g^2 
 - c (f^2 + 2 d g)) - 2 a c f g \mu + c^2 f \mu^2)
\\
& &\left.
 +  2 b d (a^3 c d + a^2 (c f - g^2) \mu + 2 a c g \mu^2 - c^2 \mu^3)
 \right), \\
\\
\langle \psi \tilde{\psi}^4 \rangle & =& q_1 \Delta^{-2}
\left(a^4 d (f^2 - d g) + 2 a^3 d (-2 b c f + b g^2 + c d \mu - f g \mu)
\right.\\
& &  +  a^2 (b^2 (3 c^2 d
 + 2 c f g - g^3) - 2 b c d g \mu
  + d (2 c f + g^2) \mu^2)
\\
& & \left.
 + c^2 \mu (2 b^3 c - b^2 g \mu + d \mu^3)
 - 2 a c (b^3 c g + b^2 (c f - g^2) \mu + d g \mu^3) \right), \\
\\
\langle \tilde{\psi}^5 \rangle & =& q_1 \Delta^{-2}
\left(-a^4 (c d^2 + f^3 - 2 d f g)
- c^2 \mu^2 (3 b^2 c - 2 b g \mu + f \mu^2)  \right. \\
& & + 2 a^3 (b f (2 c f - g^2) - (c d f - f^2 g + d g^2) \mu)\\
& &  +  a^2 (b^2 c (-5 c f + g^2) + 2 b g (-c f + g^2) \mu
 - (2 c f^2 - 4 c d g + f g^2) \mu^2)
\\
& & \left.
  +  2 a c (b^3 c^2 + b^2 c g \mu + 2 b (c f - g^2) \mu^2
 + (-c d + f g) \mu^3) \right).
\end{eqnarray*}

Six-point correlation functions in the A/2 theory are given by
\begin{eqnarray*}
\langle \psi^6 \rangle & =& q_2 \Delta^{-2} \left((a b d - b^2 g
 + b f \mu - d \mu^2) (-2 b^3 c + d \mu^3 - b \mu (3 a d + f \mu)
\right. \\
& & \left. + b^2 (2 a f + g \mu)) \right), \\
\\
\langle \psi^5 \tilde{\psi} \rangle & = & q_2 \Delta^{-2} \left(-b^5 c^2
 + a b^4 (2 c f + g^2) + a d^2 \mu^4 -  a b d \mu^2 (3 a d + 2 f \mu)
\right. \\
& & -  a b^3 (a (f^2 + 2 d g)
 + 2 (c d + f g) \mu) +  a b^2 (a^2 d^2 + 4 a d f \mu
\\
& & \left. + (f^2 + 2 d g) \mu^2) \right), \\
\\
\langle \psi^4 \tilde{\psi}^2 \rangle & = & q_2 \Delta^{-2} \left(
(b^2 c - a b f + a d \mu) (2 a^2 b d + b^2 c \mu
 + a (-2 b^2 g + b f \mu - d \mu^2)) \right), \\
\\
\langle \psi^3 \tilde{\psi}^3 \rangle & = & q_2 \Delta^{-2} 
\left(-a^4 b d^2 - a^2 b^3 (2 c f + g^2) - b^3 c^2 \mu^2 +  a b^3 c (b c + 2 g \mu)
\right. \\
& & \left. +  a^3 (b^2 (f^2 + 2 d g)
 - 2 b d f \mu + d^2 \mu^2) \right), \\
\\
\langle \psi^2 \tilde{\psi}^4 \rangle & = & q_2 \Delta^{-2} \left(
-(a^2 d - a b g
 + b c \mu) (-b c \mu^2 + a^2 (-2 b f + d \mu)
\right. \\
& & \left. +
   a b (2 b c + g \mu)) \right), \\
\\
\langle \psi \tilde{\psi}^5 \rangle & = & q_2 \Delta^{-2}
\left(a^5 d^2 - a^4 b (f^2 + 2 d g) - b c^2 \mu^4
 +  a b c \mu^2 (3 b c + 2 g \mu) \right. \\
& & +  a^3 b (b (2 c f + g^2)
  + 2 (c d + f g) \mu)
 \\
& & \left.  -  a^2 b (b^2 c^2 + 4 b c g \mu + (2 c f + g^2) \mu^2)
\right), \\
\\
\langle \tilde{\psi}^6 \rangle & = & q_2 \Delta^{-2} \left(
-(a^2 f + c \mu^2
 - a (b c + g \mu)) (2 a^3 d - c \mu^3 -
   a^2 (2 b g + f \mu)
\right. \\
& & \left.  + a \mu (3 b c + g \mu)) \right).
\end{eqnarray*}

\section{Tangent bundle moduli}
\label{app:moduli}

In this appendix we compute\footnote{
These computations were originally worked out in collaboration with
R.~Donagi and J.~Guffin for another project.
} the dimension of the tangent space to the
moduli space of tangent bundle deformations, at the tangent bundle and
`near' the tangent bundle.
We will see that
the rank of the tangent space to the moduli space
of tangent bundle deformations can change as one moves away from the (2,2)
locus.

Define
\begin{displaymath}
W \: = \: V \otimes {\cal O}(1,0) \: + \:
\tilde{V} \otimes {\cal O}(0,1),
\end{displaymath}
where
\begin{displaymath}
V \: \cong \: {\mathbb C}^{n+1}, \: \: \:
\tilde{V} \: \cong \: {\mathbb C}^{m+1},
\end{displaymath}
so that we can write the definition of the tangent bundle deformation
${\cal E}$ as
\begin{displaymath}
0 \: \longrightarrow \: {\cal O}^2 \: \longrightarrow \: W \:
\longrightarrow \: {\cal E} \: \longrightarrow \: 0.
\end{displaymath}

First, if we dualize the definition above and take the associated long
exact sequence, then from the fact that
\begin{displaymath}
H^q(W^*) \: = \: \: \: \: \mbox{ for all }q,
\end{displaymath}
(from the Bott formula, \cite{okonek}[section 1.1]), we have that
\begin{displaymath}
H^q({\cal E}^*) \: = \: H^{q-1}({\cal O}^2)
\end{displaymath}
and so vanishes unless $q=1$.

Then, applying Hom$({\cal E},-)$ to the definition of ${\cal E}$ and
taking the associated long exact sequence, one finds
\begin{displaymath}
0 \: \rightarrow \: H^0({\cal E}^* \otimes W) \: \rightarrow \:
H^0({\cal E}^* \otimes {\cal E}) \: \rightarrow \:
{\mathbb C}^4 \: \rightarrow \:
H^1({\cal E}^* \otimes W) \: \rightarrow \:
H^1({\cal E}^* \otimes {\cal E}) \: \rightarrow \: 0.
\end{displaymath}
From this expression we find 
\begin{equation}  \label{eq:bundledefs:1stpass}
h^1({\cal E}^* \otimes {\cal E}) \: = \:
h^0({\cal E}^* \otimes {\cal E}) - \left( 
h^0({\cal E}^* \otimes W) - h^1({\cal E}^* \otimes W) \right)
- 4 .
\end{equation}

Next, we will derive a relation between $h^0({\cal E}^* \otimes W)$
and $h^1({\cal E}^* \otimes W)$.
Apply Hom$(-,W)$ to the definition of ${\cal E}$ to get, from the
associated long exact sequence,
\begin{displaymath}
0 \: \rightarrow \: H^0({\cal E}^* \otimes W) \:
\rightarrow \: H^0(W^* \otimes W) \: \rightarrow \:
H^0( {\cal O}^2 \otimes W) \:
\rightarrow \: H^1( {\cal E}^* \otimes W) \:
\rightarrow \: 0,
\end{displaymath}
where we have used the fact that
\begin{displaymath}
H^q(W^* \otimes W) \: = \: 0 \: \: \: \mbox{ for }q > 0,
\end{displaymath}
as none of ${\cal O}$, ${\cal O}(1,-1)$, ${\cal O}(-1,1)$ have any
cohomology in degree greater than zero.
From the sequence above, we have that
\begin{displaymath}
h^0({\cal E}^* \otimes W) \: - \: h^1({\cal E}^* \otimes W) \: = \:
h^0(W^* \otimes W) \: - \: h^0({\cal O}^2 \otimes W) .
\end{displaymath}
To simplify further, we use the fact that
\begin{displaymath}
H^0(W^* \otimes W) \: = \:
V \otimes V^* \: + \: \tilde{V} \otimes \tilde{V}^*,
\end{displaymath}
and so has dimension
\begin{displaymath}
(n+1)^2 \: + \: (m+1)^2
\end{displaymath}
Similarly, from Bott-Borel-Weil,
\begin{displaymath}
H^0(W) \: = \: V \otimes V^* \: + \: \tilde{V} \otimes \tilde{V}^*,
\end{displaymath}
and so has the same dimension.  Thus,
\begin{displaymath}
h^0({\cal E}^* \otimes W) \: - \: h^1({\cal E}^* \otimes W) \: = \:
- (n+1)^2 - (m+1)^2.
\end{displaymath}

Plugging into equation~(\ref{eq:bundledefs:1stpass}), we find
\begin{equation}
h^1( {\rm End}\, {\cal E}) \: = \:
h^0( {\rm End}\, {\cal E}) \: + \:
(n+1)^2 \: + \: (m+1)^2 \: - \: 4.
\end{equation}

From the relation above, we immediately see that
\begin{displaymath}
h^1( {\rm End}\, {\cal E}) \: \geq \:
(n+1)^2 \: + \: (m+1)^2 \: - \: 4
\: = \: n(n+2) \: + \: m(m+2) \: - \: 2.
\end{displaymath}

Let us compute $h^0({\rm End}\, {\cal E})$ on the (2,2) locus,
where ${\cal E}$ is the tangent bundle of 
${\mathbb P}^n \times {\mathbb P}^m$.
From the Bott formula \cite{okonek}[section 1.1], one has
\begin{displaymath}
H^0({\mathbb P}^n, \Omega^1) \: = \: 0
\end{displaymath}
and from applying Hom$(T {\mathbb P}^n, -)$ to the Euler sequence,
one can similarly derive
\begin{displaymath}
h^0({\rm End}\, T {\mathbb P}^n) \: = \: h^1(\Omega^1) \: = \: 1,
\end{displaymath}
from which one quickly derives that
\begin{displaymath}
h^0({\rm End}\, {\cal E}) \: = \: 2.
\end{displaymath}

Thus, on the (2,2) locus, we find
\begin{displaymath}
h^1({\rm End}\, {\cal E}) \: = \: n(n+2) \: + \: m(m+2).
\end{displaymath}

For ${\mathbb P}^1 \times {\mathbb P}^1$, the
predicted number of infinitesimal deformations
of the tangent bundle is
$3 + 3 = 6$, matching the number of parameters on which (0,2) computations
depend, namely
\begin{displaymath}
a, \: b, \: c, \: d, \: \mu, \: \nu,
\end{displaymath}
as described in section~\ref{sect:p1p1-moduli}.

Away from the tangent bundle itself, the computations above suggest
that the correct number of moduli is smaller, which can be confirmed from
other computations.  For example, if we twist the tangent bundle
by ${\cal O}(0,-1)$, we get a rank two vector bundle of $c_2 = 2$,
and from \cite{rf}[chapter 6, theorem 20], the moduli space of such
vector bundles has dimension\footnote{
We would like to thank Z.~Lu for pointing out this computation to us.
} $4 c_2-3 = 5$.  As twisting by line bundles
does not affect Mumford stability, the space of tangent bundle
deformations should have the same dimension, so we see that the tangent
bundle of ${\mathbb P}^1 \times {\mathbb P}^1$ represents an unstable
point on the moduli space.

For higher-dimensional products, not all of the
deformations can be realized in the Euler sequence,
or as $E$ moduli in the GLSM \cite{dlm}.
For example, the predicted number of 
infinitesimal moduli of
the tangent bundle of ${\mathbb P}^1 \times {\mathbb P}^2$ is
$3 + 2(4) = 11$.
However, only seven parameters appear in the (0,2) GLSMs:
\begin{displaymath}
a, \: b, \: c, \: d, \: \mu, \: \nu_1, \: \nu_2 .
\end{displaymath}

\end{document}